\documentclass{elsart}
\usepackage{amssymb}

\renewcommand{\bar}[1]{\overline{#1}}

\usepackage{indentfirst}
\usepackage{psfig,color}
\usepackage{epsfig}
\usepackage{epsf}
\usepackage{graphicx}

\providecommand{\Journal}[4] {#1 {\bf #2} (#4) #3}
\providecommand{\EPJA}{Eur. Phys. J. A } %
\providecommand{\MPLA}{Mod. Phys. Lett. A} %
\providecommand{\NPA}{Nucl. Phys. A } %
\providecommand{\NPB}{Nucl. Phys. B } %
\providecommand{\PLB}{Phys. Lett. B } %
\providecommand{\PRL}{Phys. Rev. Lett. } %
\providecommand{\PRD}{Phys. Rev. D } %
\providecommand{\PRSA}{Proc. Roy. Soc. A } %
\providecommand{\RMP}{Rev. Mod. Phys. } %
\providecommand{\RMP}{Rev. Mod. Phys. } %
\providecommand{\ZPA}{Z. Phys. A } %

\journal{Physics Letters B}

\begin{document}

\begin{frontmatter}
\title{Pentaquark $\Theta^*$ States in the 27-plet from Chiral Soliton Models }

\author[pku]{Bin Wu},
\author[ccastpku]{Bo-Qiang Ma\corauthref{cor}}
\corauth[cor]{Corresponding author.} \ead{mabq@phy.pku.edu.cn}
\address[pku]{Department of Physics, Peking University, Beijing 100871, China}
\address[ccastpku]{CCAST (World Laboratory), P.O.~Box 8730, Beijing
100080, China\\
Department of Physics, Peking University, Beijing 100871, China}

\begin{abstract}
We estimate the mass and the width of pentaquark $\Theta^*$ states
in the 27-plet from chiral soliton models. The calculations show
that the mass of $\Theta^*$ is about 1.60~GeV and the width for
the process $\Theta^*\rightarrow \mbox{KN}$ is less than 43~MeV.
We also discuss the search for the existence of $\Theta^*$ states
in physical processes.
\end{abstract}

\end{frontmatter}

\par

Skyrme's old idea \cite{Skyr} that baryons are solitons has been
widely accepted since Witten's topological analysis of the
Wess-Zumino term and his clarification in what sense baryons can
be considered as classical solitons of effective meson fields
\cite{witt}. The quantization of the SU(3) Skyrmion not only gives
the baryon octet and decuplet, but also predicts new higher baryon
multiplets, such as the anti-decuplet, the 27-plet etc
\cite{Guad,Mano,Chem}. There are exotic baryon states with
strangeness number $S=+1$ in these higher multiplets, and these
states can be interpreted as pentaquark states with minimal
five-quark configurations uuud$\bar{s}$, uudd$\bar{s}$, and
uddd$\bar{s}$ in the quark language \cite{GM99}. A number of
authors \cite{Pra,penta1,Diak,Weig} predicted the mass of the
lightest pentaquark $\Theta^+(uudd\bar{s})$ state with S=+1 from
chiral soliton models. However, the real boost in searching
pentaquark states was due to Diakonov, Petrov and Polyakov's
prediction about the mass and the width of $\Theta^+$ \cite{Diak}.
It seems that recent experiments \cite{LEPS,DIAN,CLAS,SAPH,HERMES}
have revealed the existence of $\Theta^+$ with a mass
$M_{\Theta^+}\simeq$1.54~GeV, $S=+1$ and a very small width
$\Gamma_{\Theta^+}<$25~MeV. From the absence of a signal in the
corresponding pK$^+$ invariant mass distribution in $\gamma p
\rightarrow pK^+K^-$ and $\gamma^*  p\rightarrow pK^+K^-$ at the
expected strength \cite{SAPH,HERMES}, it is suggested that
$\Theta^+$ should be an isoscalar. Thus up to now, experiments
have given a surprising support to prediction from chiral soliton
models. In Ref.~\cite{Wall}, Walliser and Kopeliovich studied the
other exotic states in the 27-plet and the $\overline{35}$-plet
and predicted the mass of the higher $\Theta$ states in the
27-plet, called $\Theta^*$ in the following, to be about
1.65/1.69~GeV, provided that the mass of $\Theta^+$ is at
1.54~GeV. Borisyuk, Faber, and Kobushkin \cite{BFK03} also
predicted the mass of $\Theta^*$ around $1595$~MeV and the width
at 80~MeV by identifying N(1710) as the member of the
anti-decuplet.

\par

In this letter, we calculate both the mass and the width of
pentaquark $\Theta^*$ states from chiral soliton models, and give
the predictions based on available experimental observations. The
motivation for this is to reveal the existence of pentaquark
$\Theta^*$ states through further experiments, following along the
successful prediction from Ref.~\cite{Diak}.

\par

Following Ref. \cite{Guad}, the SU(3) symmetric effective action
in the large $N_c$ limit leads to the collective Hamiltonian:
\begin{equation}
    \widehat{H}=M_{cl}+\frac{1}{2I_2}\left[\widehat{C}^{(2)}-\frac{1}{12}(N_cB)^2\right]+\left(\frac{1}{2I_1}-\frac{1}{2I_2}\right)\mathbf{\widehat{J}}^2,
\end{equation}
where $M_{cl}$ is the classical soliton mass;
$\widehat{C}^{(2)}$=$\sum\limits_{a=1}^{8}\widehat{G}_a^2$ is the
quadratic (Casimir) operator of the vectorial group SU(3)$_v$, in
the representation $(p,q)$, its eigenvalue
$C^{(2)}=\frac{1}{3}[p^2+q^2+pq+3(p+q)]$; $\widehat{G}_a~~(a=1-8)$
are the generators of SU(3)$_v$; $\widehat{J}_i~~(i=1-3)$ are the
the generators of the spin group SU(2)$_s$; $I_1$ and $I_2$ are
moments of inertia. Therefore, for the representation $(p,q)$ of
the SU(3)$_v$ and the spin $J$, the eigenvalues of the Hamiltonian
are
\begin{equation}
E^{(p,q)}_J=M_{cl}+\frac{1}{6I_2}\left[p^2+q^2+pq+3(p+q)-\frac{1}{4}(N_cB)^2\right]
+\left(\frac{1}{2I_1}-\frac{1}{2I_2}\right)J(J+1).
\end{equation}
From the energy eigenvalues above, it can be argued that the
27-plets with spin 3/2 and 1/2 are the multiplets next to the
antidecuplet \cite{Chem}. The mass differences between the lowest
multiplets are
\begin{eqnarray}\nonumber
    &&(1,1)-(0,3): E^{(8)}-E^{(\overline{10})}=-\frac{3}{2I_2},\\\nonumber
    &&(1,1)-(2,2): E^{(8)}-E_{\frac{1}{2}}^{(27)}=-\frac{5}{2I_2},\\\nonumber
    &&\\\nonumber
    &&(2,2)-(1,4): E^{(27)}_\frac{3}{2}-E^{(\overline{35})}_\frac{3}{2}=-\frac{2}{I_2},\\\nonumber
    &&(2,2)-(3,3): E^{(27)}_\frac{3}{2}-E_{\frac{3}{2}}^{(64)}=-\frac{7}{2I_2},\\\nonumber
\end{eqnarray}
\par

The states of the system will correspond to the baryon states, and
wave function $\Psi_{\nu\nu^\prime}^{(\mu)}$ of baryon $B$ in the
collective coordinates is of the form
\begin{equation}
    \Psi_{\nu\nu^\prime}^{(\mu)}(A)=\sqrt{\mbox{dim}(\mu)}D^{(\mu)}_{\nu\nu^\prime}(A),A \in
SU(3),
\end{equation}
where $(\mu)$ denotes an irreducible representation of the SU(3)
group; $\nu$ and $\nu^{\prime}$ denote $(Y, I, I_3)$ and $(1, J,
-J_3)$ quantum numbers collectively; $Y$ is the hypercharge of
$B$; $I$ and $I_3$ are the isospin and its third component of $B$
respectively; $J_3$ is the third component of spin $J$;
$D^{(\mu)}_{\nu\nu^\prime}(A)$ are representation matrices.
However, due to the non-zero strange quark mass, the
 symmetry breaking Hamiltonian is \cite{Bolt}
\begin{eqnarray}
    &&H^\prime=\alpha D^{(8)}_{88}+\beta
    Y+\frac{\gamma}{\sqrt{3}}\sum_{i=1}\limits^3D^{(8)}_{8i}J^i\label{Hp},
\end{eqnarray}
where the coefficients $\alpha$, $\beta$, $\gamma$ are
proportional to the strange quark mass and model dependent, but
they are treated model-independently and fixed by experiments in
this letter; $D^{(8)}_{ma}(A )$ is the adjoint representation of
the SU(3) group and defined as:
\begin{equation}
D^{(8)}_{ma}(A
)=\frac{1}{2}\mbox{Tr}(A^{\dagger}\lambda^mA\lambda^a),
\end{equation}
and $\lambda^m$ is the Gell-Mann matrix of the corresponding
meson.
\par Another consequence of the flavor symmetry breaking is
that a physical baryon state is no long a pure state belonging to
a unique multiplet, but a mixing state with the corresponding
members with identical spin and isospin in other multiplets, that
is \
\begin{equation}
    \Psi_{\nu\nu^\prime}(A)=\sum\limits_\mu c_{\nu\nu^\prime}^{(\mu)}
    \Psi^{(\mu)}_{\nu\nu^\prime}(A).
\end{equation}
From (\ref{Hp}), the physical baryon states are of the form by
first-order approximation
\begin{eqnarray}\nonumber
        &&\left|N\right>=\left|N;8\right>+C_{\overline{10}}
        \left|N;\overline{10}\right>+C_{27}\left|N;27_\frac{1}{2}\right>,\\\nonumber
        &&\left|\Theta^{+}\right>=\left|\Theta^{+};\bar{10}\right>,\\\nonumber
        &&\left|\Theta^{*}\right>=\left|\Theta^{*};27_\frac{3}{2}\right>
        +C_{\bar{35}}\left|\Theta^{*};\bar{35}_\frac{3}{2}\right>+C_{64}\left|\Theta^{*};64_\frac{3}{2}\right>.
\end{eqnarray}
To linear order of $m_s$, the coefficients above are given simply
by perturbation theory
\begin{eqnarray}
    \begin{array}{ll}
    C_{\overline{10}}=-\frac{1}{3\sqrt{5}}(\alpha+
    \frac{\gamma}{2})I_2, &C_{27}=-\frac{\sqrt{6}}{25}(\alpha-
    \frac{\gamma}{6})I_2.\\
    C_{\overline{35}}=-\frac{3}{4\sqrt{35}}\left(\alpha+\frac{5}{6}\gamma\right)I_2,&
    C_{64}=-\frac{3\sqrt{10}}{196}\left(\alpha-\frac{1}{6}\gamma\right)I_2.
    \end{array}
\end{eqnarray}
In chiral soliton model, the 27-plet with spin $\frac{3}{2}$,
lower than that with spin 1/2, is the next multiplet to the
anti-decuplet, we only deal with spin-3/2 baryons in this letter,
and omit the spin-3/2 index of the notations of particles in the
27-plet as well as energy eigenvalue from now on. The quark
content of the exotic pentaquark states are suggested in Fig.~1,
and the mass splittings of the isomultiplets in the 27-plet are
listed in Table 1.\\\par

\begin{tabular}{|l|l|l|l|l|}
\multicolumn{5}{c}{Table 1. The mass of baryons in the \{27\}
multiplet}
\\ \hline Baryon &I&Y&$\left<B|H^\prime|B\right>$&Mass~(GeV)\\ \hline exotic pentaquarks&&&&\\
\hline
$\Theta^{*}$&1&2&$\frac{\alpha}{7}+2\beta-\frac{5}{14}\gamma$&1.60\\
$X_{1s}$&2&0&$\frac{5}{56}\alpha-\frac{25}{112}\gamma$&1.68\\
$X_{2s}$&$\frac{3}{2}$&-1&$-\frac{1}{14}\alpha-\beta+\frac{5}{28}\gamma$&1.87\\
$\Omega^{*}$&1&-2&$-\frac{13}{56}\alpha-2\beta+\frac{65}{112}\gamma$&2.07\\
\hline
$\Delta^{*}$&$\frac{3}{2}$&1&$\frac{13}{112}\alpha+\beta-\frac{65}{224}\gamma$&1.64\\
\hline exited states of octet&&&&\\ \hline
$N_{27}$&$\frac{1}{2}$&1&$\frac{1}{28}\alpha+\beta-\frac{5}{56}\gamma$&1.73\\
$\Sigma_{27}$&1&0&$-\frac{1}{56}\alpha+\frac{5}{112}\gamma$&1.80\\
$\Xi_{27}$&$\frac{1}{2}$&-1&$-\frac{17}{112}\alpha-\beta+\frac{85}{224}\gamma$&1.96\\
$\Lambda_{27}$&$0$&0&$-\frac{1}{14}\alpha+\frac{5}{28}\gamma$&1.86\\
\hline
\end{tabular}

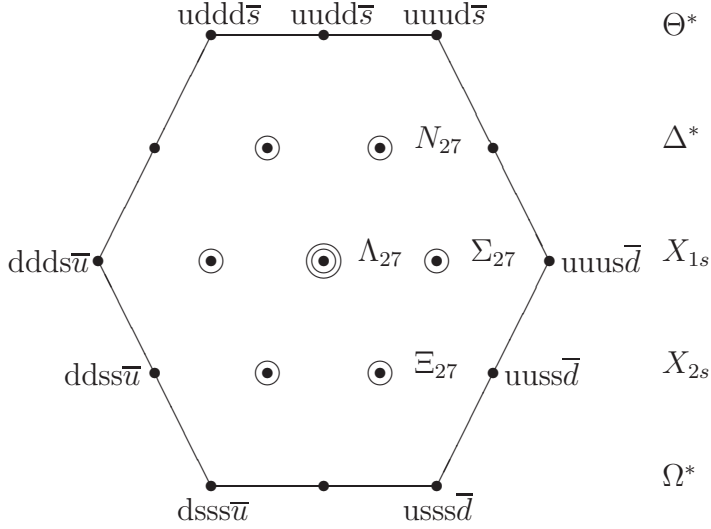
\begin{figure}
\setlength{\unitlength}{1.5cm}
\begin{picture}(5.5,5.5)(-1.5,-1)
\put(1,0){\line(1,0){2}}\put(1,0){\line(-1,2){1}}
\put(0,2){\line(1,2){1}} \put(1,4){\line(1,0){2}}
\put(3,4){\line(1,-2){1}} \put(4,2){\line(-1,-2){1}}

\put(1,0){\circle*{0.1}}\put(2,0){\circle*{0.1}}\put(3,0){\circle*{0.1}}
\put(0,2){\circle*{0.1}}\put(1,4){\circle*{0.1}}\put(2,4){\circle*{0.1}}
\put(3,4){\circle*{0.1}}\put(4,2){\circle*{0.1}}\put(3.5,1){\circle*{0.1}}
\put(3.5,3){\circle*{0.1}}\put(0.5,1){\circle*{0.1}}\put(0.5,3){\circle*{0.1}}

\put(0.7,-0.3){dsss$\overline{u}$}
\put(2.7,-0.3){usss$\overline{d}$}\put(-0.8,1.9){ddds$\overline{u}$}
\put(0.7,4.1){uddd$\overline{s}$}\put(1.7,4.1){uudd$\overline{s}$}\put(2.7,4.1){uuud$\overline{s}$}
\put(4.1,1.9){uuus$\overline{d}$}\put(3.6,0.9){uuss$\overline{d}$}\put(-0.3,0.9){ddss$\overline{u}$}
\put(5,0){$\Omega^{*}$}\put(5,1){$X_{2s}$}\put(5,2){$X_{1s}$}
\put(5,3){$\Delta^{*}$}\put(5,4){$\Theta^*$}

\put(1.5,1){\circle*{0.1}}\put(1.5,3){\circle*{0.1}}\put(1,2){\circle*{0.1}}
\put(2.5,3){\circle*{0.1}}\put(2.5,1){\circle*{0.1}}\put(3,2){\circle*{0.1}}
\put(2,2){\circle*{0.1}}
\put(2.8,1){$\Xi_{27}$}\put(2.8,3){$N_{27}$}\put(3.3,2){$\Sigma_{27}$}

\put(1.5,1){\circle{0.2}}\put(1.5,3){\circle{0.2}}\put(1,2){\circle{0.2}}
\put(2.5,3){\circle{0.2}}\put(2.5,1){\circle{0.2}}\put(3,2){\circle{0.2}}
\put(2,2){\circle{0.2}}\put(2.3,2){$\Lambda_{27}$}
\put(2,2){\circle{0.3}}
\end{picture}
\caption{The quark content of the \{27\} multiplet baryons.}
\end{figure}\label{fig1}

\par
In experiments, we are interested in the decay
$\Theta^*\rightarrow KN$ which are realized by a pseudoscalar
Yukawa coupling. In soliton models, such a coupling can be
obtained by Goldberger-Treiman relation, which relates the
relevant coupling constant to the axial charge \cite{Adki,Blot1}.
And up to $1/N_c$ order, the coupling operator in the space of the
collective coordinates $A$ has the form \cite{Diak,Blot1}:
\begin{equation}
    \widehat{g}_A\propto~G_0D^{(8)}_{m3}
    -G_1d_{3ab}D^{(8)}_{ma}J_b
    -\frac{G_2}{\sqrt{3}}D^{(8)}_{m8}J_3,
\end{equation}
where $d_{iab}$ is the SU(3) symmetric tensor, $a, b=4,5,6,7$, and
$J_a$ are the the generators of the infinitesimal SU$_R$(3)
rotations. $G_{1}$, $G_{2}$ are dimensionless constants, $1/N_c$
suppressed relative to $G_0$. Let $\left|i\right>$
=$\Psi^{(\mu^\prime)}_{\rho\rho^\prime}(A)+
c_{i}\Psi^{(\mu_i)}_{\rho\rho^\prime}(A)$ denote the state of
$B^\prime$ and $\left|f\right>$ =$\Psi^{(\mu)}_{\nu\nu^\prime}(A)+
d_{f} \Psi^{(\mu_f)}_{\nu\nu^\prime}(A)$ denote the state of $B$,
where $c_{i}$ and $d_{f}$ are chosen to be real and are all of
1/$N_c$ order. Then sandwiching $\widehat{g}_A$ between
$\left|f\right>$ and $\left|i\right>$ gives the coupling
$g_{BB^\prime m}$ for the decay $B \rightarrow B^\prime m$
\begin{eqnarray}
g^2_{BB^\prime
m}=\frac{g_0}{(2J_{\nu}+1)}\sum\limits_{J_{\nu3}J_{\rho3}}\sum\limits_{I_{m3}I_{\rho3}}\left|\left<f\right|G_0D^{(8)}_{m3}
    -G_1d_{3ab}D^{(8)}_{ma}J_b
    -\frac{G_2}{\sqrt{3}}D^{(8)}_{m8}J_3\left|i\right>\right|^2,~~~~
\end{eqnarray}
where we leave $g_0$ as a constant to be fixed by experiments; and
it is averaged over the initial spin($J_{\nu},J_{\nu3}$) and sums
over the final spin($J_{\rho},J_{\rho3}$) as well as
isospin($I_{m},I_{m3}$ and $I_{\rho},I_{\rho3}$) states. Up to
linear order of $m_s$ and 1/$N_c$ and neglecting the terms
proportional to $m_s/N_c$, we can rewrite the coupling as
\begin{eqnarray}\nonumber
    \lefteqn{g^2_{BB^\prime
m}}\\\nonumber
    &&=\frac{g_0G^2}{3}\times\left\{\begin{array}{c}
    \frac{\mbox{dim}(\mu^\prime)}{\mbox{dim}(\mu)}\left|\begin{array}{c}\sum\limits_\gamma\left(
    \begin{array}{cc}8&\mu^\prime\\ Y_mI_m&Y_\rho I_\rho \end{array}\right|
    \left.\begin{array}{c}\mu_\gamma\\Y_\nu
    I_\nu\end{array}\right)
    \left(
    \begin{array}{cc}8&\mu^\prime\\01&1 J_\rho \end{array}\right|
    \left.\begin{array}{c}\mu_\gamma\\1J_\nu\end{array}\right)\end{array}\right|^2+\end{array}\right.\\\nonumber
    &&\left.\begin{array}{c}
    \\2c_i\frac{G_0}{G}\frac{\sqrt{\mbox{dim}(\mu^\prime)\mbox{dim}(\mu_i)}}{\mbox{dim}(\mu)}\begin{array}{c}\sum\limits_\gamma\left(
    \begin{array}{cc}8&\mu^\prime\\Y_mI_m &Y_\rho I_\rho \end{array}\right|
    \left.\begin{array}{c}\mu_\gamma\\Y_\nu
    I_\nu\end{array}\right)
    \left(
    \begin{array}{cc}8&\mu^\prime\\01&1 J_\rho\end{array}\right|
    \left.\begin{array}{c}\mu_\gamma\\1J_\nu\end{array}\right)\end{array}\end{array}\right.\\\nonumber
    &&\left.\begin{array}{c}
    \\\times\begin{array}{c}\sum\limits_{\gamma^\prime}\left(
    \begin{array}{cc}8&\mu_i\\Y_mI_m &Y_\rho I_\rho \end{array}\right|
    \left.\begin{array}{c}\mu_{\gamma\prime}\\Y_\nu
    I_\nu\end{array}\right)
    \left(
    \begin{array}{cc}8&\mu_i\\01 &1J_\rho \end{array}\right|
    \left.\begin{array}{c}\mu_{\gamma^\prime}\\1J_\nu\end{array}\right)\end{array}+\end{array}\right.\\\nonumber
    &&\left.\begin{array}{c}
    \\2d_f\frac{G_0}{G}\frac{\mbox{dim}(\mu^\prime)}{\sqrt{\mbox{dim}(\mu)\mbox{dim}(\mu_f)}}\begin{array}{c}\sum\limits_{\gamma}\left(
    \begin{array}{cc}8&\mu^\prime\\Y_mI_m &Y_\rho I_\rho \end{array}\right|
    \left.\begin{array}{c}\mu_{\gamma}\\Y_\nu
    I_\nu\end{array}\right)
    \left(
    \begin{array}{cc}8&\mu^\prime\\01 &1J_\rho \end{array}\right|
    \left.\begin{array}{c}\mu_{\gamma}\\1J_\nu\end{array}\right)\end{array}\end{array}\right.\\\nonumber
    &&\left.\begin{array}{c}
    \\\times\begin{array}{c}\sum\limits_{\gamma_i}\left(
    \begin{array}{cc}8&\mu^\prime\\Y_mI_m &Y_\rho I_\rho \end{array}\right|
    \left.\begin{array}{c}\mu_{f\gamma_i}\\Y_\nu
    I_\nu\end{array}\right)
    \left(
    \begin{array}{cc}8&\mu^\prime\\01 &1J_\rho \end{array}\right|
    \left.\begin{array}{c}\mu_{f\gamma_i}\\1J_\nu\end{array}\right)\end{array}\end{array}\right\},\\
\end{eqnarray}
where $G$ can be extracted from $\left<f\right|G_0D^{(8)}_{m3}
    -G_1d_{3ab}D^{(8)}_{ma}J_b
    -\frac{G_2}{\sqrt{3}}D^{(8)}_{m8}J_3\left|i\right>$ \cite{Diak}.
    In this approximation, we ignore such cases that some of the SU(3) flavor
Clebsch-Gordan coefficients which would multiply $G_{1,2}$ so that
would have $N_{c}$ dependence enhancing the naive $N_{c}$ power.
This is what happens in the case of
$G_{\overline{10}}=G_{0}-\frac{N_{c}+1}{4}G_{1}- \frac{1}{2}G_{2}$
where the constant $G_{1}$, which is formally $O(1/N_{c})$ with
respect to $G_{0}$, is enhanced \cite{pras}. The discussion of
this effect is beyond the scope of the present paper and an
improved width formula is used to discuss the width of the
anti-decuplet baryons in Ref.~\cite{aapsw}. Then, the width is
given by
\begin{eqnarray}\nonumber
    \Gamma(B\rightarrow B^\prime m)=\frac{3g^2_{BB^\prime
m}}{4\pi
m_B}|\mathbf{p}|\left[(m_{B^\prime}^2+\mathbf{p}^2)^{\frac{1}{2}}-m_{B^\prime}\right]\approx\frac{3g^2_{BB^\prime
m}}{8\pi m_Bm_B^\prime}|\mathbf{p}|^3.\\\nonumber
\end{eqnarray}

This formula is the same as that in Refs.~\cite{Diak} and
\cite{Weig} in the non-relativistic case. It is very well
satisfied experimentally in the case of the decay of the decuplet
baryons, the input data \cite{PDG} and the numeric results are as
listed in Table 2.

\begin{tabular}{|l|lll|}
\multicolumn{4}{c}{Table 2. The best fit at $g_0$=3.84}\\\hline
Decay modes&$m_B$,$m_B^\prime,m_m~\mbox{(MeV)}$ & PDG data &theory
values\\\hline
$\Delta\rightarrow N\pi$ &1232,983.3,139.6&$\approx$120& 117\\
$\Sigma^*\rightarrow\Lambda\pi$ &
1385,1116,135&$\approx$34.67&34.4\\
$\Sigma^*\rightarrow\Sigma\pi$&1387,1197,135&$\approx$4.73& 4.5\\
$\Xi^*\rightarrow\Xi\pi$&1535,1321,135&$\approx$9.9&10.4\\\hline
\end{tabular}
\\
\\

After a trivial calculation, we have the width formulae for
$\Theta$ and $\Theta^*$ decays:
\begin{eqnarray}
    \Gamma(\Theta^+ \rightarrow KN)=g_0\frac{(G_0-G_1-\frac{1}{2}G_2)^2}
    {40\pi m_{\Theta^+}m_N}|\mathbf{p}|^3
    [1+\frac{G_0}{G_0-G_1-\frac{1}{2}G_2}(\frac{\sqrt{5}}{2}C_{\overline{10}}-
    \frac{7\sqrt{6}}{12}C_{27})],~~~~~~\\
    \Gamma(\Theta^* \rightarrow KN)=g_0\frac{(G_0-\frac{1}{2}G_1)^2}
    {54\pi m_{\Theta^*}m_N}|\mathbf{p}|^3
    [1-\frac{G_0}{G_0-\frac{1}{2}G_1}(\frac{\sqrt{5}}{2}C_{\overline{10}}+
    \frac{3\sqrt{6}}{28}C_{27})].~~~~~~\\\nonumber
\end{eqnarray}

Ref.~\cite{NA49} first reported evidence for the existence of a
narrow $\Xi^-\pi^-$ baryon resonance with mass of
$1.862\pm0.003$~GeV and width below the detector resolution of
about $0.018$~GeV, and this state is considered as a candidate for
the pentaquark $\Xi^{--}_{\frac{3}{2}}$. If we take both
$\Theta^+$ and the candidate for $\Xi_{3/2}$ \cite{NA49} as
members of the anti-decuplet and solve the following equations
\begin{equation}\left\{\begin{array}{l}\nonumber
\frac{1}{I_1}=\frac{2}{3}\left[E^{(10)}-E^{(8)}\right]=\frac{2}{3}\left[m_{\Sigma^*}
-\frac{1}{2}(m_\Lambda+m_\Sigma)\right]=154~\mbox{MeV};\\\nonumber
\alpha+\frac{3}{2}\gamma=5(m_\Lambda-m_\Sigma)=-385~\mbox{MeV};\\\nonumber
\frac{1}{8}\alpha+\beta-\frac{5}{16}\gamma=m_\Delta-m_{\Sigma^*}=-153~\mbox{MeV};\\\nonumber
E^{(\overline{10})}+(\frac{1}{4}\alpha+2\beta-\frac{1}{8}\gamma)=m_{\Theta^+}=1540~\mbox{MeV};\\\nonumber
E^{(\overline{10})}-(\frac{1}{8}\alpha-\beta+\frac{1}{16}\gamma)=m_{\Xi_{3/2}}=1860~\mbox{MeV};\\\nonumber
E^{(\overline{10})}-\frac{3}{2I_2}=E^{(10)}-\frac{3}{2I_1}=1154.5~\mbox{MeV};\\\nonumber
E^{(27)}-E^{(\overline{10})}+\frac{1}{2I_2}=\frac{3}{2I_1}=230.5~\mbox{MeV};
\end{array}\right.
\end{equation}

we get

\[
\begin{array}{llll}
m_{\Theta^*}=1.60~\mbox{GeV};&E^{(27)}=1.785~\mbox{GeV};&
1/I_2=399~\mbox{MeV};&\\ \alpha=-663~\mbox{MeV};&
\beta=-12~\mbox{MeV}; & \gamma=185~\mbox{MeV};&\\
C_{\overline{10}}=0.21;&
C_{27}=0.17;&C_{\bar{35}}=0.16;&C_{35}=0.08.
\end{array}\]

We find that the values of $C_{\overline{10}}$ and $C_{27}$ are
the same as those in Ref.~\cite{Lee}. The masses of baryons for
the 27-plet in the parameters above are listed in Table 1. Using
the results above, we can calculate the width for
$\Theta^*\rightarrow KN$:
\begin{equation}
    \Gamma(\Theta^* \rightarrow KN)=g_0\frac{(G_0-\frac{1}{2}G_1)^2}
    {54\pi m_\Theta^*m_N}|\mathbf{p}|^3
    [1-\frac{G_0}{G_0-\frac{1}{2}G_1}(\frac{\sqrt{5}}{2}C_{\overline{10}}+
    \frac{3\sqrt{6}}{28}C_{27})]\leq43~\mbox{MeV}.
\end{equation}
Thus, soliton model gives a stringent restriction on the width of
$\Theta^*$ for the process $\Theta^*\rightarrow KN$, and if, as
reported by recent experiments, $\Gamma_{\Theta^+}<25~\mbox{MeV}$,
the width for $\Theta^*\rightarrow KN$ will be less than 43~MeV.
The predicted width could be more narrow if a smaller input width
$\Gamma_{\Theta^+}$ is used.

In the pentaquark $\Theta^*$ triplet, $\Theta^{*++}$ and
$\Theta^{*+}$ may be easily measured. The search for $\Theta^{*+}$
is similar to $\Theta^{+}$ through the decay modes $\Theta^{*+}
\to K^+ n$ and $K^0 p$ with approximately same magnitudes
\cite{Chen}. There have been suggestions for search of pentaquark
$\Theta^{++}(uuud\bar{s})$ state in virtual and real photon
scattering on the proton target \cite{GM99,Cap}. Provided with the
ranges of the predicted mass and width, the existence of
$\Theta^{*++}$ may be revealed through the decay mode
$\Theta^{*++}\to K^+ p$ in various processes, such as:\\
Photon-nucleon collisions\\
$\gamma p\rightarrow\Theta^{*++}K^-;~\Theta^{*++}\rightarrow
pK^+$;\\
Nucleon-nucleon collisions\\
$pp\rightarrow pK^-\Theta^{*++};~\Theta^{*++}\rightarrow pK^+$;\\
Pion-nucleon collisions\\
$\pi^+ p \rightarrow \bar{K}^0
\Theta^{*++};~\Theta^{*++}\rightarrow
pK^+$;\\
Electron(virtual photon)-nucleon collisions\\
$e p\rightarrow e'K^-\Theta^{*++};~\Theta^{*++}\rightarrow
pK^+$.\\
$\Theta^{*0}$ may be revealed through the decay mode
$\Theta^{*0}\to K^0 n$ in the above corresponding processes with
the target changed from proton to neutron. The correlations
between the constructed $K N$ invariant masses of $\Theta^{*++}$,
$\Theta^{*+}$, and $\Theta^{*0}$ decays can test whether the
corresponding states are belong to the pentaquark $\Theta^{*}$
triplet suggested in this work.

In summary, calculations from the chiral soliton model show that,
the pentaquark $\Theta^{*}$ states in the 27-plet with spin 3/2,
have a mass around 1.60~GeV and a width for $\Theta^*\rightarrow
KN$ less than 43~MeV. The existence of these pentaquark $\Theta^*$
states can be revealed by the decay modes $\Theta^* \to K N$ in
various physical processes.

{\bf Acknowledgments }

We are grateful for discussions with Yanjun Mao, Han-Qing Zheng
and Zhi-Yong Zhao. This work is partially supported by National
Natural Science Foundation of China under Grant Numbers 10025523
and 90103007.


\end{document}